# CONDITOR[1]: Topic Maps and DITA labelling tool for textual documents with historical information

Piedad Garrido[1], Jesús Tramullas[2], Manuel Coll[1]

[1] Department of Computer Science, Universidad de Zaragoza,
44003 Teruel, Spain
[2] Department of Information Science, Universidad de Zaragoza,
55009 Zaragoza, Spain
{piedad, ,tramullas}@unizar.es, mkhollv@gmail.com

**Abstract.** Conditor is a software tool which works with textual documents containing historical information. The purpose of this work two-fold: firstly to show the validity of the developed engine to correctly identify and label the entities of the universe of discourse with a labelled-combined XTM-DITA model. Secondly to explain the improvements achieved in the information retrieval process thanks to the use of an object-oriented database (JPOX) as well as its integration into the Lucene-type database search process to not only accomplish more accurate searches, but to also help the future development of a recommender system [21]. We finish with a brief demo in a 3D-graph of the results of the aforementioned search.

**Keywords:** Topic Maps, XTM, DITA, JPOX, Lucene, object-oriented databases, historical information, cultural heritage.

## 1 Introduction

Textual documents containing historical information are a required source of reference for socio-cultural and historical research development. Besides, their importance grows exponentially as they are increasingly used in learning systems and virtual learning environments, and in the diffusion of the digital environment culture which addresses all audience types. The success of the Perseus Digital Library Project (http://www.perseus.tufts.edu) is a clear example of this. The growing number of historical document collections, both textual and multimedia, available at digital libraries will demand the development of more advanced data access tools and information presentation tools where Topic Maps will play a central role.

One of the main features of textual documents with historical information is their poor level of data structure. The wide range of document types that can be used in history research shows a lack of easy definable patterns, and when they do exist, they correspond to a highly specific series of documents. Then there is the additional problem of adding the irregular appearance of significant historical entities (people's names, locations, dates, acts, etc.), and what this entails within the textual documents themselves. Identifying these entities is one of the problems we face when we wish to integrate these documents into semantic web environments. On numerous occasions, such problems have been resolved by using specially built ontologies as a reference, for instance, those derived from the conceptual model CIDOC CRM (http://cidoc.ics.forth.gr/) for museum documentation and cultural heritage contexts. On other occasions, automatic identification techniques of entities and relationships have been used. Folch and Habert (2000) proposed the mining-text methods to extract significant semantically terms, which have been automatically labelled later to generate Topic Maps directly. Recently, Robertson and Dicheva (2007) proposed a tool to generate a topic map draft, which must subsequently be edited and validated by the user.

Given their potential application, from the very beginning documents about cultural heritage have been the object of the research community with regard to Topic Maps (Sigel 2000). Nevertheless, we appreciate that most of the available bibliography insists on issues of relationships among elements, and on matters of integration into the presentation and editing of Topic Maps tools. Combining the structured labelling of documents with Topic Maps has already been proposed on different occasions, for example, the Topic Maps integration in TEI (Vasallo, 2005). Other researchers have opted to develop structures from the document content itself by establishing relationships manually (Nagase and Naito, 2002).

---

1   A minor Roman God, whose specialized work was storage.

Obviously, the main problem of using Topic Maps in historical documents lies in the identification and labelling of the significant semantic elements, and in the establishment of the relationships among them. The manual approach sometimes used through the edition and direct labelling of the document content entails having to use numerous human resources as well as the possibility of mistakes being made. The aim of this work is to show the validity of the software tool developed to correctly identify and label entities as a previous step to Topic Map generation and to obtain visual representations of the universe of discourse by applying heuristics capable of processing the original document and of transforming it properly.

## 2 Topic Maps-DITA Approach

The project in which this work is framed started out as the development of a end-user recommender system of the GEA on-line product (Gran Enciclopedia Aragonesa) which helps the user during information retrieval processes. Initially, the aim actually became a specialized processing tool of cultural heritage information.

```
− <voces>
  − <voz subcategoriaId="38">
      <vozId> 98 </vozId>
      <nombre> Abd al-Malik ibn Hudayl ibn Razin </nombre>
    − <descripcion>
        <p>Segundo soberano de la $$taifa%%$$ de Albarracín, entre 1045 y 1103, con el título de Hu
        al-Dawla (Sable del Estado). Fue muy criticado por Ibn Hayyan, historiador contemporáneo suyo, a
        Ja-qa-n le pondera con ditirambos. Pero su tiempo, al volcar el siglo xi con la decadencia de las taifa
        permite ya la altiva independencia de los $$Banu%%$$ Razin, sus antepasados. En 1085, después d
        VI tomara Toledo, sintiendo el peligro cerca le envió sumiso sus felicitaciones por la conquista. En
        tributaba al Cid diez mil dinares, pero cuando lo vio lejos sitiando Valencia, dejó de pagárselos, into
        aliarse con el rey de Aragón; éste avisó al Cid, antes de agosto de 1093, quien vino a atacarle y le r
        vez. Pero cuando Rodrigo ganó Valencia, el señor de Albarracín se alió con los $$almorávides%%79
        lado peleó en la batalla del Cuarte, abandonando el sitio en vergonzosa fuga. Murió en la $$Sahla%
        mayo de 1103.</P>
      </descripcion>
  </voz>
  − <voz subcategoriaId="38">
      <vozId> 99 </vozId>
      <nombre> Abd al-Rahman I </nombre>
    − <descripcion>
        <p>Primer emir $$omeya%%$$ de Al-Andalus (757-788). Proclamado con apoyo, entre otros, d
        $$yemeníes%%13105$$, éstos mayoría en la Marca Superior se le alzaron pronto y allí envía a sus g
        Tammam ibn Alqama (en 764) y Badr (en 767), que deportó a Córdoba a $$Sulayman%%$$ al-Arabi.
        rebela, en Zaragoza, Suwayd ibn Musà; contra él fue el emir en persona, antes de derrotar en el Be
        una coalición yemení, cuya venganza asume Sulayman al-Arabi, propagando la sedición por todo el
```

Fig. 1. Example of a text input register

The information provided by the on-line encyclopedia managers did not offer the minimum semantic labelling (see figure 1) required for the development of the end-user tool proposed. The absence of semantic, for example, is evident and the voice field <description> has to be analyzed more deeply because there is a lot of important information dependent on the context such as events, dates, places, people, marriages, descendants, etc.

Previously, a reprocessing of the above-mentioned information has had to be performed. The decision to work with superimposed information (Tramullas and Garrido, 2006) to combine XTM and DITA was taken. As this was a case of textual historical information management, the use of these labelling languages has enabled, as far as XTM is concerned, the creation of a conceptual structure design (see figure 2) and future maintenance for editing, browsing and combining the voices coupled with support for relating concepts, linking concepts to resources, merging ontologies, external searching for resources, defining perspectives, etc. On the other hand, DITA allows for the possibility to obtain a contender which takes the best out of DocBook (http://www.docbook.org) or Information Mapping (http://www.infomap.com), or other information architectures, and also allows for a more flexible presentation of information. And, it is firmly rooted in open source ideals.

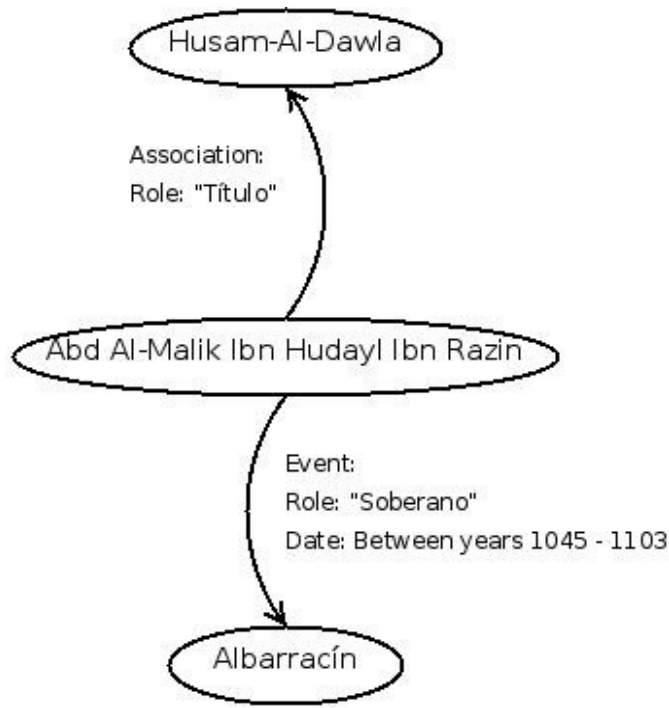

**Fig. 2. Example of the conceptual structure proposed**

Once the textual documents have been reprocessed with this approach, the labelling obtained has to be enriched semantically (see example):

```xml
<voces xmlns:ditaarch="http://dita.oasis-open.org/architecture/2005/" xmlns:xlink="http://www.w3.org/1999/xlink" >
<topic id="98">
    <baseName>
        <baseNameString>Abd al-Malik ibn Hudayl ibn Razin</baseNameString>
        <variant>
            <variantName>
                <resourceData>Abd al-Malik ibn Hudayl ibn Razin</resourceData>
            </variantName>
        </variant>
    </baseName>
    <instanceOf>
        <topicRef xlink:type="simple" xlink:show="replace" xlink:actuate="onRequest" xlink:href="#38"/>
    </instanceOf>
    <contents>
        <shortdesc>
            Segundo soberano de la taifa de Albarracín, entre 1045 y 1103, con el título de Husam Al-Dawla (
        </shortdesc>
        <body>
            Segundo soberano de la taifa de Albarracín, entre 1045 y 1103, con el título de Husam al-Dawla (

        </body>
    </contents>
    <date>
        <role>soberano</role>
        <location>Albarracín</location>
        <year>1045</year>
    </date>
    <date>
        <role>soberano</role>
        <location>Albarracín</location>
        <year>1103</year>
    </date>
    <date>
        <role>murió</role>
        <location>Sahla</location>
        <day>18</day>
        <month>5</month>
        <year>1103</year>
    </date>
    <occurrence>
            <roleSpec>soberano</roleSpec>
        <resourceData>Albarracín</resourceData>
    </occurrence>
    <occurrence>
```

**Fig. 3. XTM-DITA output file**

Once the semantically enriched information is available, the engineering software process, which develops an application to work with this collection of voices, is clearly favored. Other authors like Drobnik and Ueberall (2006) have demonstrated this. This developmental part of the research and its corresponding prototype software, whose central element is Conditor which we now go on to explain, have allowed, among other things, a simple XML document of a poor semantic description to be converted into an XML document with a structure based on a Topic Maps-DITA combination.

## 3 CONDITOR Engine Development

### 3.1 Introduction

The main objective of the Conditor program consists in: reading the XML source (see figure 1), which in our case is an XML file format; carrying out the generation of entities which represents each of the source structural elements (objects) during this reading stage (see figure 2); generating an XML object with a different structure from these entities (see figure 3); and finally, enabling their storage in any system.
Naturally, this multifactor aim leads us to think of numerous technologies that provide an automatic conversion among XML documents, as well as the automatic generation of objects from sources in this format. Nevertheless, the problem left to be solved involves the possibility that the XML document takes a complex structure which reflects semantic richness. In this case, a simple conversion tool cannot undertake such a process (Roberson and Dicheva, 2007).

If we go into this situation more deeply, we find that the root of the problem lies in extracting the semantic data required in the target text from the original source to achieve a properly performed conversion. In other words, the process must enrich the information during conversion. The kind of historical information being managed in this research project contains entity names, that is, people's names, locations, etc. Entity names, just as Bikel, Schwartz and Weischedel shows in one of his works (1999), offer a whole range of possibilities irrespectively of the fact that they are classified into broad categories or that they are used in an accurate classification, as they may prove very useful in information retrieval processes which use natural language processing (NLP) (Cerny, 2006), question answering and the automated construction of ontologies. Fleischman and Hovy (2002) highlights that: "while locations can often be classified based solely on the words that surround the instance, person names are often more challenging because classification relies on much deeper semantic intuitions gained from the surrounding text". So, part of our research into semantic work has centered on the subcategorization of people's names and what this process entails.

Furthermore, it is necessary for the enrichment to be stored in the objects generated during the reading in order to make full use of the additional information generated in all the storage systems, thus achieving a highly improved retrieval process of subsequent searches. In this particular case, the program must convert a simple XML document with poor semantic aspects into an XML document whose structure is based on the combination of DITA and XML Topic Maps (XTM). Other aspects the program must accomplish are the storage of information in an object-oriented database, JPOX, and the indexing of objects text in a Lucene database (see figure 4).

### 3.2 Conditor Engine

The purpose of this program written in Java is to meet the aims proposed in the previous section. To do this, several processes may be used to provide various reading methods, interpretation, writing and storage.

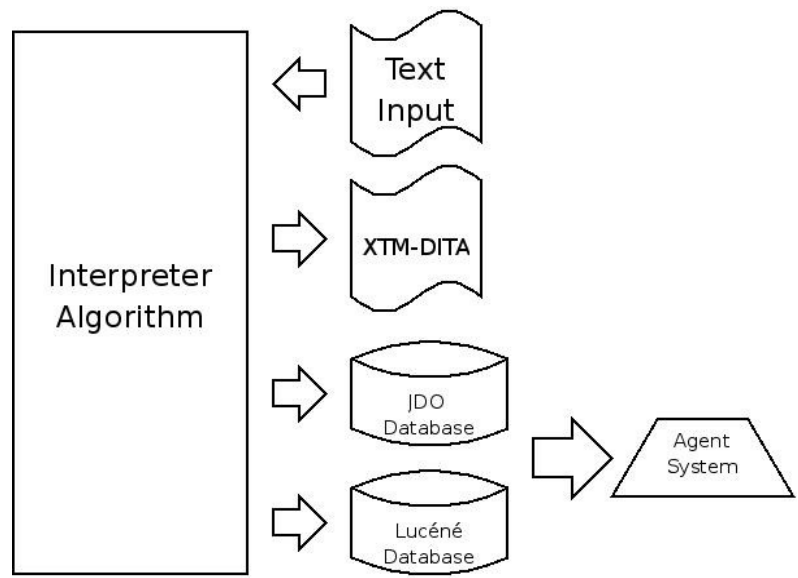

**Fig. 4.** System Arquitecture Diagram

The processing sequence that the program must follow is as so (see figure 4):

- Initialize and start the reading of the introduced XML source (text input).
- Generate the objects which correspond to the source data.
- Interpret the body of each entry of the source document to fill the generated objects.
- Start the XML document generation to the XTM-DITA format from the data of the generated objects.
- Establish the persistence of these objects by means of an object-oriented database (JDO Database).
- Index the textual content of the objects in a Lucene database.
- Offer support to do searches on Internet about a related topic via an agent system.

The program will open the document and will use a reprogrammed XML analyzer to generate the necessary objects for each XML label in a way that is automatic and clear. Each object generated corresponds to a source document entry, and only fields directly associated with the XML field of the source are filled at this time. Subsequently (and also simultaneously to the previous step), the system is provided with a new series of rules, algorithms or mechanisms that are alien to it which permit the following operations:

- Full-Text reprocessing: analysis algorithm of the entry title, detecting and merging names, text cleaning, organizing texts in words and/or phrases. This reprocessing is necessary to improve the effectiveness of the algorithm which operates subsequently and works with clean structured texts.
- Interpreting the body of the entry: algorithms that detect patterns of dates, events, places, people, instruments or roles associated with the entry itself. These algorithms are the pillar of semantic enrichment information since they extract new data, establish their classification and reflect them in the new XTM-DITA target structure (see figure 3).
- Detection of occurrences, one-way or two-way relationships with the rest of the entries: crossing-search algorithms that are able to find occurrences between topics (see figure 2) and which establish one-way and two-way associations, and may establish the role of the associations through textual analysis algorithms.

.
All these proposed algorithms must be modular so that they can be externally modified by other people. The purpose of this characteristic is to refine these mechanisms so that they specialize in specific domains, thus providing certain versatility and to improve them with new techniques without having to recompile the whole system.

The algorithms used to interpret the entry body start from simple patron-matching algorithms based on word-by-word reading. They are also based on the use of ontology databases where semantic relationships among words are established which pass through natural language interpreters able to obtain and catalog information, which are in turn based on complete phrase patterns, until learning systems founded on contextual relationships among words are established which are able to classify upon the appearance of keywords in phrases. However, all these text extraction approaches have their advantages and drawbacks.

The systems based on word-by-word reading are efficient in very specific domains, for instance, the people's names entries. These systems are capable of separating aliases from names, and of extracting information of dates and associations based on names (Fleischman, 2001). The use of ontology databases, which semantically check word by word and which establish a classification, is also possible (Vargas-Vera et all., 2000). Although this system is primitive, the information gained through it more than justifies its application.

The systems which use phrase recognition have to be founded on natural language interpretation systems in such a way that they are capable of extracting and classifying information based on specific patterns. One example of their application can be observed in (Lui, Chi and Ng, 2003). Finally, learning systems must be able to find context aspects both inside and outside phrases which are based on the frequency of key words appearing, and which indicate contexts and discover relationships among words.

Our approach is based on the use of heuristics designed for this field of application. Working with these heuristics in an intelligent agent environment (see figure 3), which performs searches at an internal and external level of application, helps to provide a high level of independence which will allow this application to be used in various fields of knowledge. This approach has also verified the reliability of using XTM-DITA combined as proxies for semantic content. Finally, we wish to point out that the semantic information extraction has been worked by parsing an XML document obtained through a search performed by a user. Once object enrichment has been accomplished by means of entries description interpretations, we proceed to the final phase: target document generation, and storing and indexing all the information in the corresponding databases. Target document generation is a direct process where each object is mapped with an XML document entry so that all the objects of each field match each XML entry section.

Next, object persistence is carried out where the objects in the object-oriented database (JPOX) are stored. The advantage of using JPOX as opposed to other approaches which make use of relational database management systems, like those described in (Schmidt and Stephan, 2006), consists in indicating what objects and what fields are persistent via a simple XML description, and in establishing their representation in the database. More importantly, however, is that JPOX is an object-oriented database from the programmer's point of view and that it performs an automatic conversion so that the final data are stored through the simulation of a relational database repository. This has to be a two-way and transparent conversion for users. Then, this database will have to enable this resulting database repository to be used in order to extract entries from its repository identifier. Finally, all the words of all the object descriptions generated are indexed in a Lucene database (LeVan's, 2003), which relates all the words that appear in an entry with the identifier to subsequently use this database to perform complex word searches to obtain the related entry identifier, and to obtain the information contained in the repository database from this identifier. The Lucene integration used in information retrieval processes where Topic Maps are used may be seen at Horvati (2006).

## 4. Conclusions and Future Developments

In this paper, we advocated several ideas: that good information processing is essential, especially in areas with poor levels of information structure and in Spanish language; that modularity in software design is extremely important; and that combining object-oriented databases, mature software applications such as Lucene APIs and Latent Semantic Indexing (LSI), etc. is extremely useful in information retrieval processes. And if all of these ideas are integrated into an intelligent agent framework, their potential is multiplied.

Modularity will allow the *Conditor* engine to run in any knowledge area without the need to recompile the whole system or to adapt it to different environments. Intelligent agents (see figure 4) offer many advantages in information retrieval processes and leave the system ready for the future development of recommender systems. At the same time, they play a fundamental role in graphical user interface generation.

The object-oriented database management system, JPOX, and the Lucene database enable XTM-DITA tags to be suitable managed because the relationships among objects are more flexible and enriched in comparison with other systems developed under relational database management systems (QuaaxTM-A).

Even though previous developments concerning historical culture management have been carried out in this research field, such works focused on information retrieval processes to facilitate navigation and subsequent presentations, as seen in the citations included throughout the text and in the bibliography provided. *Conditor* covers the whole life cycle of this information treatment since it reprocesses, labels, indexes and finally generates a visual representation, as seen in Figure 6.

**Fig.5** Current Graphical User Interface

.

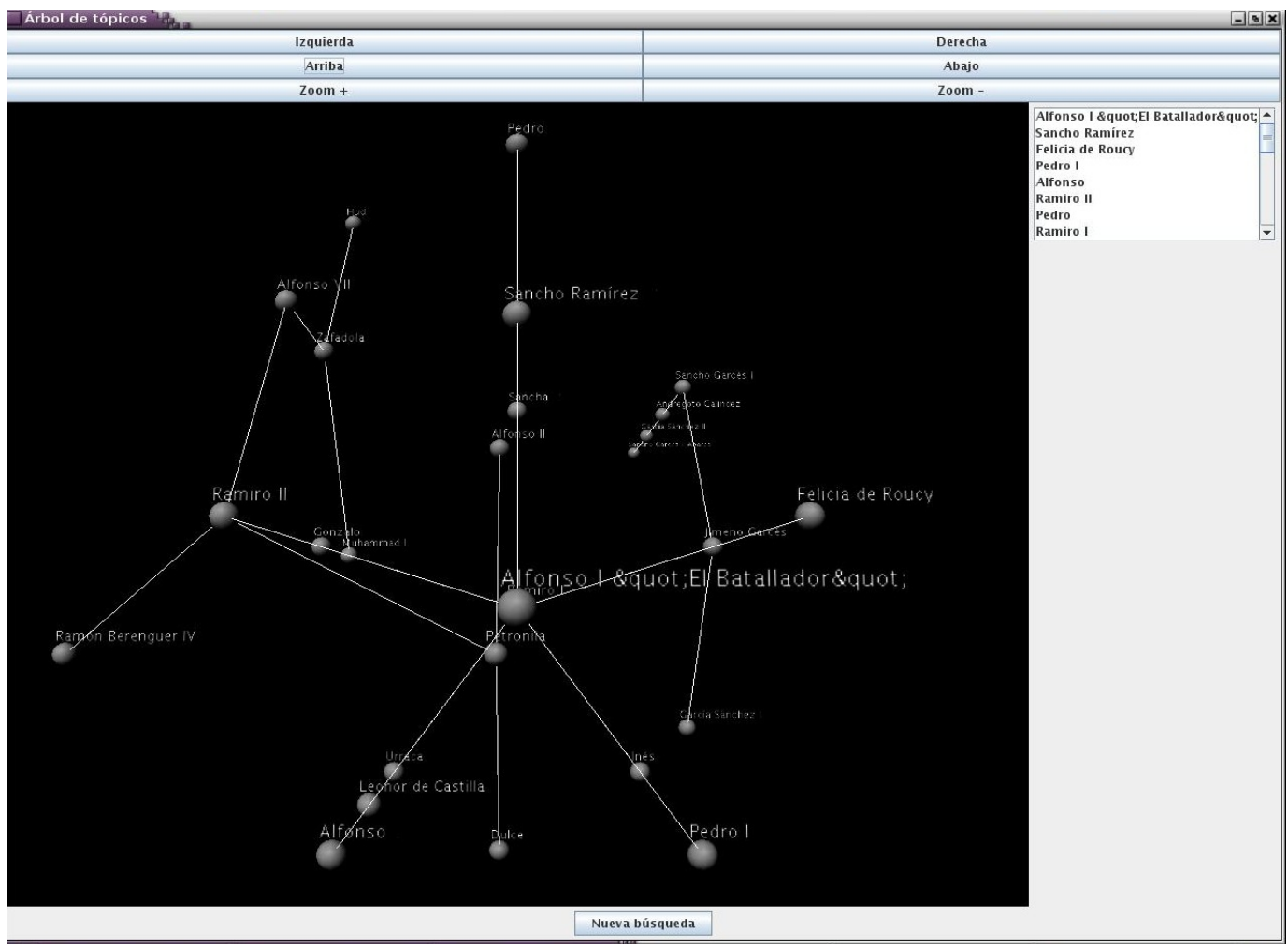

**Fig. 6** Graphical User Interface Prototype